\documentclass[amsmath,twocolumn,prl,aps]{revtex4-1}
\usepackage{amsthm,amsfonts,bm,graphicx,verbatim,color}
\usepackage{subcaption,float,dcolumn}
\usepackage[utf8]{inputenc}

\newcommand{\p}{\partial}
\newcommand{\beqn}{\begin{eqnarray}}
\newcommand{\eeqn}{\end{eqnarray}}

\newcommand{\ee}{\epsilon}
\newcommand{\ohalfnu}{$\nu$=1/2 }
\newcommand{\quartnu}{$\nu$=1/4 }
\newcommand{\thirdnu}{$\nu$=1/3 }
\newcommand{\fcflnu}{$\nu$=$1/2m$ }
\newcommand{\afcflnu}{$\nu$=1$-1/2m$ }

\DeclareMathOperator{\sign}{sgn}
\DeclareMathOperator{\Tr}{Tr}

\begin{document}
\title{Dirac Fermion Hierarchy of Composite Fermi Liquids}

\author{Jie Wang}
\email{jiew@princeton.edu; jiewang.phy@gmail.com.}
\affiliation{Department of Physics, Princeton University, Princeton, New Jersey 08544, USA}

\begin{abstract}
Composite Fermi liquids (CFLs) are compressible states that can occur for 2D interacting fermions confined in the lowest Landau level at certain Landau level fillings. They have been understood as Fermi seas formed by composite fermions which are bound states of electromagnetic fluxes and electrons as reported by Halperin, Lee and Read [Phys. Rev. B 47, 7312 (1993)]. At half filling, an explicitly particle-hole symmetric theory based on Dirac fermions was proposed by Son [Phys. Rev. X 5, 031027 (2015)] as an alternative low energy description. In this work, we investigate the Berry curvature of CFL model wave functions at a filling fraction one-quarter, and observe that it is uniformly distributed over the Fermi sea except at the center where an additional $\pi$ phase was found. Motivated by this, we propose an effective theory which generalizes Son's half filling theory, by internal gauge flux attachment, to all filling fractions in which fermionic CFLs can occur. The numerical results support the idea of internal gauge flux attachment.
\end{abstract}

\maketitle
Composite Fermi liquids (CFLs) are gapless states that can occur at certain Landau level fillings $\nu$. They were first explained by Halperin, Lee, and Read (HLR) \cite{HalperinLeeRead} as Fermi seas (FSs) of electromagnetic flux attached composite fermions. While it succeeded in explaining CFL's metallic feature, it is not obvious how the HLR theory is consistent with the particle-hole (PH) symmetry \cite{WangHalperinPH,Mulliganwhyhlr,DunghaiLeeNeutralFermions}, which is an exact symmetry in a half filled Landau level if Landau level mixing is negligible. Recently, an emergent Dirac fermion (DF) theory was proposed by Son \cite{Son} at half filling. With $\psi$ as the DF field, $\gamma^{\mu}$ as the gamma matrix, $a_{\mu}$ and $A_{\mu}$ as internal and external gauge fields respectively, Son's DF action is
\beqn
\mathcal{L} = i\bar\psi\gamma^{\mu}\left(\p_{\mu} - ia_{\mu}\right)\psi - \frac{1}{2}\frac{1}{2\pi}adA + \frac{1}{2}\frac{1}{4\pi}AdA. \label{DiracFermionAction}
\eeqn
where $e,\hbar$ are set to be 1, so the magnetic length is $l_B^{-2}$=$B_A$, with $B_A$=$\ee^{ab}\p_aA_b$ as the external magnetic field strength and $\ee^{xy}$=$-\ee^{yx}$=1 as the antisymmetric symbol. Greek and latin letters label space-time and spatial coordinates respectively. Higher order terms in Eq.~(\ref{DiracFermionAction}) are omitted for simplicity. Son's DF theory is explicitly PH symmetric because PH acts in a way akin to time reversal on DFs. It also suggests intriguing dualities between Dirac and nonrelativistic fermions in two dimensions \cite{Son,dualityweb,tongduality,2ddualityMotrunich,WireConstructionMulligan,MaxHFLL,WangSenthil,ThermoAshvin,ScottPHBoson}.

Son's half filled DF theory predicts a $\pi$ Berry phase, which in fact is a $\pi$ Berry curvature singularity located at the FS center (as DF is a two-component spinor), acquired by the composite fermion (DF) when transported around the Fermi surface. This $\pi$ Berry phase and Berry curvature have been observed from numerics \cite{GeraedtsScience,scottjiehaldane,Jie_MonteCarlo,Simons}. The FS Berry phase $\Phi_{FS}$ \footnote{$\Phi_{FS}$ is defined as the Berry phase acquired by the composite fermion when it is ``anticlockwisely'' transported along the Fermi surface.} has been argued to be closely tied to electron Hall conductivity $\sigma_H$ ($\sigma^{CF}_H$: composite fermion Hall conductivity): as pointed out by Haldane in the theory of anomalous Hall effect \cite{haldaneanomaloushall,NiuAnomalousHall}, the nonquantized part of Hall conductivity is determined by the FS Berry phase; see Eq.~(\ref{FSBP}). Variants of the HLR theory with an emphasis on FS Berry phase have been studied in Refs.~\cite{WangSenthilCFL,WangSenthil,yizhiyouhfll}. 
\beqn
\sigma_H^{CF} = -\sigma_H = \frac{e^2}{h}\frac{\Phi_{FS}}{2\pi},\quad \Phi_{FS}=-2\pi\nu. \label{FSBP}
\eeqn

In principle, CFLs can occur as long as the HLR flux attached particles are fermions; whether or not they occur depends on the details of interaction. When the underlying physical particles are fermions, the filling fractions of CFLs can be grouped into two classes: \fcflnu if FSs are formed by composite fermions (we denote them as fCFLs), and \afcflnu if formed by composite holes (anti-fCFLs). In this work, we studied the Berry curvature of the \quartnu model wave function (MWF) as a case study, and proposed an effective theory for fCFLs and anti-fCFLs at \emph{all} filling fractions that they can occur. This theory can be viewed as generalizing Son's DF theory by attaching each DF with $\pm|2m-2|$ internal gauge flux quanta. PH conjugate states are realized by attaching the same amount but an opposite direction of fluxes. As a result at long wavelength, CFLs for fermions can be considered as descending from the \ohalfnu PH symmetric states, in analogy to Jain's hierarchy \cite{JainCF89} which interpreted incompressible quantum Hall fluids as descending from integer quantum Hall effects.


\begin{figure}[]
    \centering
    \includegraphics[width=0.4\textwidth]{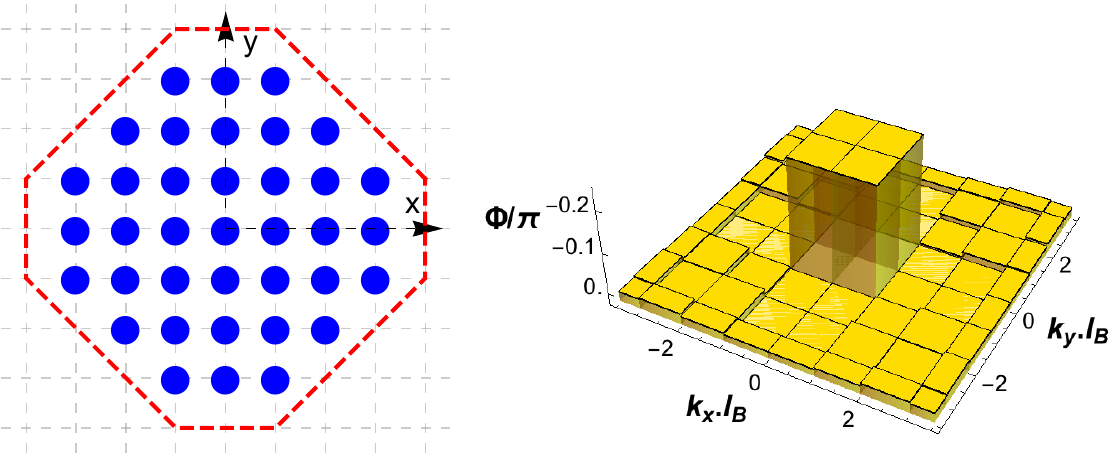}
    \captionsetup{justification=RaggedRight, singlelinecheck=true}
    \caption{Berry curvature distribution (right) obtained from $\Psi^{1/4}_{n=1}$ model wave function by a linear regression on a FS (left) consisting of $N$=37 dipoles. The red dashed line represents the path, which can be interpreted as FS boundary, along which \emph{anticlockwisely} transporting a single composite fermion has $-2\pi\nu$ Berry phase. The area enclosed by the red dashed line contains $46$ grids. The Berry curvature has a peak of around $-$0.25+$0.011$=$-$0.239 (in units of $\pi$), while the rest of the values are around $-$(2/4$-$1)/46=0.011. It suggests an interesting Berry curvature distribution for CFLs at a generic filling fraction in the thermodynamic limit: a $-\pi$ singularity at center and $-(2\pi\nu$$-$$\pi)$ uniformly distributed over FS.
    }\label{berrycurvature}
\end{figure}

Based on the numerical study on \quartnu MWFs, we found that (i) the FS contains $-2\pi\nu$ Berry phase in agreement with Eq.~(\ref{FSBP}) and (ii) the $-2\pi\nu$ FS Berry phase consists of a $-\pi$ peak located at the FS center, and a $-2\pi\left(\nu-\frac{1}{2}\right)$ phase uniformly distributed over the FS. 

Motivated by this, we conjectured an effective field theory, dubbed the \emph{flux-attached DF theory}, as follows \footnote{The same action was considered from a different perspective in \cite{FradkinOneQuarter}.}:
\beqn
\mathcal{L}&=&i\bar\psi\gamma^{\mu}\left(\p_{\mu}-ia_{\mu}\right)\psi - \frac{1}{2m}\frac{1}{2\pi}adA \label{flux-dirac-action-new}\\
&&-\eta\left(\frac{1}{2}-\frac{1}{2m}\right)\frac{1}{4\pi}ada + \left(\frac{1}{2}-\frac{\eta}{2}\frac{m-1}{m}\right)\frac{1}{4\pi}AdA.\nonumber
\eeqn
Extra prescription: an extra $-\left(2\pi\nu-\pi\right)$ Berry phase uniformly distributed over the FS.

Equation~(\ref{flux-dirac-action-new}) represents a Fermi liquid theory at generic filling fractions: \fcflnu ($\eta$=+1, fCFL) and its PH conjugation \afcflnu ($\eta$=$-$1, anti-fCFL), where $m$ is a positive integer. We conjecture that DFs are massless particles even at $m$$\neq$1 as constrained by the FS Berry phase. Equation~(\ref{flux-dirac-action-new}) is determined by the Luttinger theorem, Hall conductivity, and FS Berry phase. We will argue that Berry curvature obtained from HLR motivated wave functions agrees with the prediction out of flux-attached DF theory. Further testings of Eq.~(\ref{flux-dirac-action-new}) can be obtained from studying response functions, which we will show somewhere else.



\textbf{\emph{Model wave function.---}}
In the following, we examine CFL MWFs at \quartnu as a case study. The MWFs were proposed based on the ideas of HLR's flux attachment \cite{RezayiReadCFL,RezayiHaldaneCFL}. Key ingredients of the MWFs at $\nu$=$1/2m$ include flux attachment represented by the Jastrow factor and the lowest Landau level (LLL) projection $P_{LLL}$ operator,
\beqn
\Psi\left(\{\bm k\}, \{z\}\right) = P_{LLL}\{\det_{ia}e^{i\bm k_a\cdot \bm r_i}\prod_{i<j}^{N}(z_i-z_j)^{2m}\}.\label{HLRmwf}
\eeqn
where $\{\bm k\}$ are distinct and clustered to form a \emph{compact FS}. Holomorphic determinant MWFs are obtained after approximating \cite{JainKamilla,PuJain1,PuJain2} $P_{LLL}$ by creating dipoles $\{d_i\}$, in accordance with the dipole-momentum locking which is a fundamental property of composite fermions in a LLL. With $\sigma(z)$ as the modified Weierstrass sigma function \cite{haldanemodularinv,haldaneholomorphic}, $\{\alpha_k\}$ as the center of mass zeros which set the topological sector, MWF at $\nu$=$1/2m$ reads \cite{shaoprl},
\beqn
\Psi^{1/2m}_n(\{\bm d\},\{\alpha\},\{z\})&=&\det_{ia} M_{ia}\prod_{i<j}^{N}\sigma^{2(m-n)}(z_i-z_j)\label{modelwavefunction}\\
&&\times\prod_k^{2m}\sigma\left(\sum_i^Nz_i-\alpha_k\right)\prod_i^{N}e^{-(1/2)z_iz_i^*}.\nonumber
\eeqn
where $M_{ia}^{}$=$e^{(n/2m)z_id_a^*}\prod_{k \neq i}^{N}\sigma^n\left(z_{i}-z_{k}-d_a+\bar d\right)$. The $\bar d$ is a free parameter; {\it i.e.} changing $\bar d$ only renormalizes the MWF \cite{scottjiehaldane}. $n$ represents a scheme of flux attachment: $2n$ out of the total $2m$ flux quanta are shifted from the electron's position to form a dipole. Like momentum quantization, dipoles $\{d_i\}$ are quantized by the periodic boundary condition \cite{haldanetorus1,haldanetorus2} to take discrete values $d\in\{\mathbb{L}/(nN)\}$ where $\mathbb{L}$ is the 2D periodic lattice defining the torus.

We adopt the lattice Monte Carlo method \cite{Jie_MonteCarlo} to study the Berry phase. We consider \quartnu MWFs with $m$$\geq$$n$ (see Supplemental Material): $\Psi^{1/4}_{n=1}$ and $\Psi^{1/4}_{n=2}$. They are found to have large overlaps with each other for all dipole configurations, {\it e.g.}, $|\langle\Psi^{1/4}_{n=1}|\Psi^{1/4}_{n=2}\rangle|\geq 97\%$ for $N$=69 dipoles. This means that observables computed from either of them are almost identical.

\begin{figure}[]
    \centering
    \hspace*{-0.025\textwidth}\includegraphics[width=0.4\textwidth]{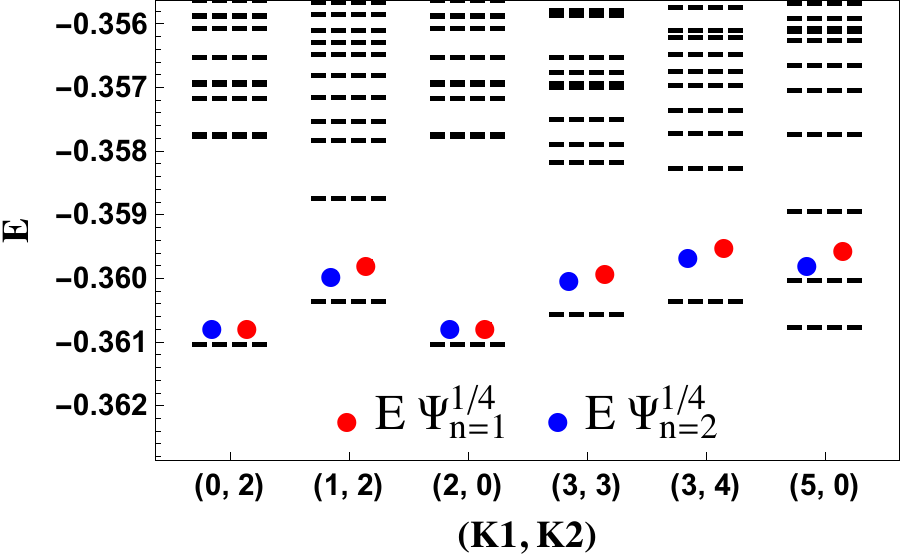}
    \captionsetup{justification=RaggedRight, singlelinecheck=true}
    \caption{Variational energies (red dots for $\Psi^{1/4}_{n=1}$, blue dots for $\Psi^{1/4}_{n=2}$) and exactly diagonalized Coulomb energies (dashed lines) as a function of many-body momentum \cite{haldanetorus2} $(K_1, K_2)$ for $N$=10 electrons for the \quartnu filled LLL on a square torus. Energies are plotted in units of $e^2/\epsilon l_B$. For each $K_1$, $K_2$ is chosen to match the momentum of the lowest energy state. Because of inversion symmetry, only $K_1\in[0, 5]$ are plotted.}\label{comparewithed}
\end{figure}

In a half filled LLL, Coulomb interaction low energy states were found to have a remarkably large overlap \cite{scottjiehaldane} with the clusterlike ansatz of Eq.~(\ref{modelwavefunction}). At one-quarter, second quantizing a MWF to compute overlap becomes difficult for large system sizes. Instead, as shown in Fig.~\ref{comparewithed}, we present the energy spectrum of the LLL Coulomb interaction and the variational energy of MWF for $N$=10 electrons on a square torus. The variational energies of MWFs and exact diagonalization energies are close, but slightly worse compared to one-half states. As pointed out in Ref.~\cite{RezayiHaldaneCFL}, at half filling in the lowest two Landau levels, varying short-range interactions induce a first-order phase transition from the striped phase to a strongly paired Moore-Read state, followed by a possible crossover to a weak paring phase. The exact diagonalization states we obtained at Coulomb point at one quarter, presumably, are weakly paired states; tuning $v_{1,3}$ pseudopotentials might help improve the overlaps. See Supplemental Material for comments for the $(5, 0)$ sector.

\begin{figure}[]
    \centering
    \hspace*{-0.025\textwidth}\includegraphics[width=0.4\textwidth]{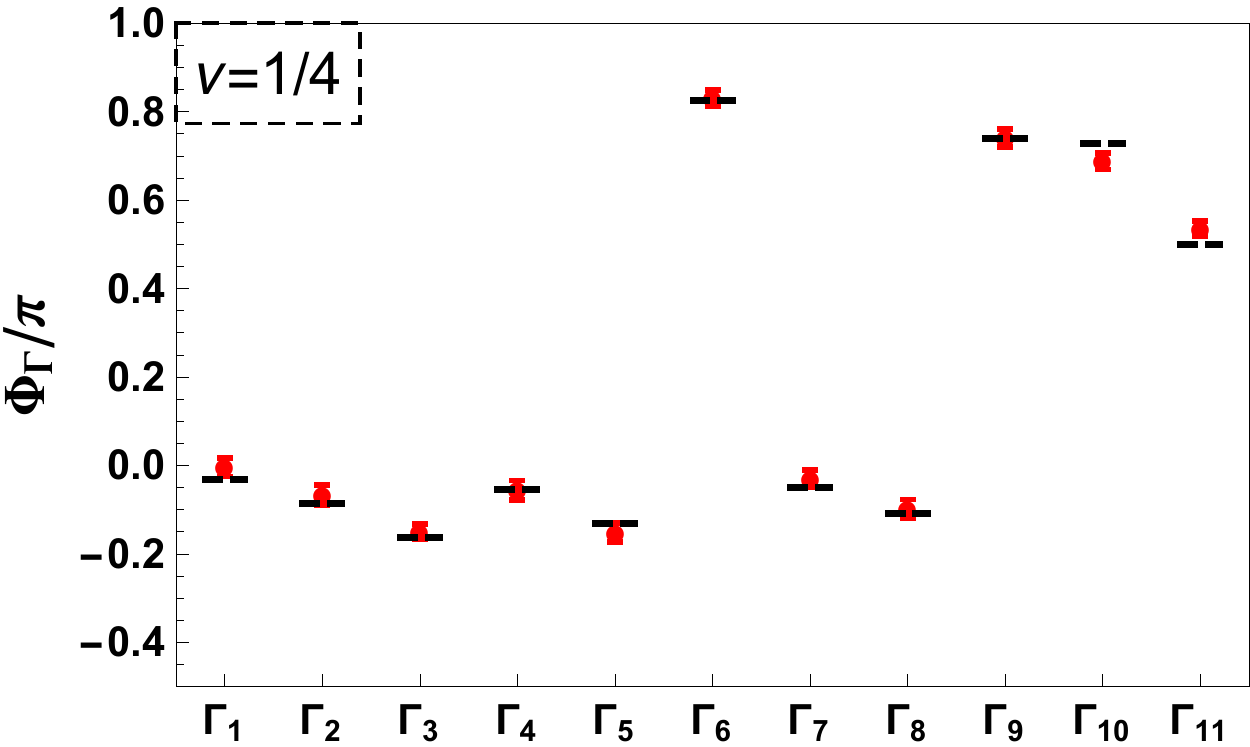}
    \captionsetup{justification=RaggedRight, singlelinecheck=true}
    \caption{Comparison of the Berry phases $\Phi_{\Gamma}$ associated with various clock-wised paths $\{\Gamma\}$=$\Gamma_1, ..., \Gamma_6$ on a FS of $N$=37 dipoles (see Fig.~\ref{berrycurvature} for FS) computed from $\Psi^{1/4}_{n=1}$ (red dots) and from the formula $\Phi_{\Gamma}$=$\delta_{\Gamma}$$\cdot$$\pi$+$(2\pi\nu$$-$$\pi)A_{\Gamma}/A_{FS}$ [black lines], where $\delta_{\Gamma}$ is the winding number of $\Gamma$ relative to the FS center, $A_{\Gamma}$ and $A_{FS}$ are the $\bm k$-space areas enclosed by the path $\Gamma$ and FS area respectively. See appendix for details about paths and more examples including $N$=69 FS and $\nu$=1/3.
    }\label{berrycurvature_37_4}
\end{figure}

\textbf{\emph{Berry curvature.---}}
We next turn to the numerical investigation of the Berry curvature shown in Fig.~\ref{berrycurvature}. Since only MWFs with compact dipole configurations are identified as CFLs \cite{scottjiehaldane,Jie_MonteCarlo,Simons}, we neither take off composite fermions deep inside the FS nor excite them too far away from the Fermi surface. Instead, Berry phases were computed on clockwised paths close to the Fermi surface, after which the Berry curvatures were mapped out by a linear regression \footnote{See Supplementary Material for further details. We found that the many-body Berry phase has an unphysical path-dependent phase factor $(i)^{N_+-N_-}$ at generic filling, which we justify is needed to make the many-body Berry phase transform consistently under PH and path orientation inversion. Additionally, the $-\pi$ of total $-2\pi\nu$ FS Berry phase is attributed to the Dirac point, and the rest $-(2\pi\nu-\pi)$ phase is an extra prescription to the Dirac Fermi sea to cancel the effects of the CS term.}. The Berry curvature distribution was found as described in Fig.~\ref{berrycurvature}. See Fig.~\ref{berrycurvature_37_4} for a consistency check which indicates that Fig.~\ref{berrycurvature} makes sense.

The same results were found even for bosonic CFL \thirdnu states \cite{pasquirehaldane,ReadLLL}. From a wave function point of view, the Berry curvature feature can be argued as follows \cite{HaldanePrivate}: the determinant (of two fluxes) and the Jastrow factor are implementing, respectively, the $\pi$ and the uniform part of the Berry phase. We believe that the Berry curvature feature we observed on \quartnu and \thirdnu model states applies to other filling fractions as well since MWFs at different Landau level fillings essentially differ only by a different power of Jastrow factors. See Supplemental Material for further discussions.

\textbf{\emph{Effective action.---}}
The presence of $\pi$ Berry curvature singularity at $\nu$$\neq$1/2 strongly suggests the emergence of DFs at low energy at generic filling fractions. In this section, we justify our proposed flux-attached DF theory Eq.~(\ref{flux-dirac-action-new}) by starting from a Dirac-type effective action with undetermined coefficients, Eq.~(\ref{undeterminedaction}). We will then fix $C_{1,2,3}$ by physical requirements: Luttinger theorem, FS Berry phase and Hall conductivity, and argue that the Berry curvature distribution is consistent with the prediction from the flux-attached DF picture.
\beqn
\hspace*{-0.5cm}\mathcal{L}=i\bar\psi\gamma^{\mu}\left(\p_{\mu}-ia_{\mu}\right)\psi - \frac{C_1}{2\pi}adA - \frac{C_2}{4\pi}ada + \frac{C_3}{4\pi}AdA.\label{undeterminedaction}
\eeqn

\begin{figure}[]
    \centering
    \hspace*{-0.025\textwidth}\includegraphics[width=0.4\textwidth]{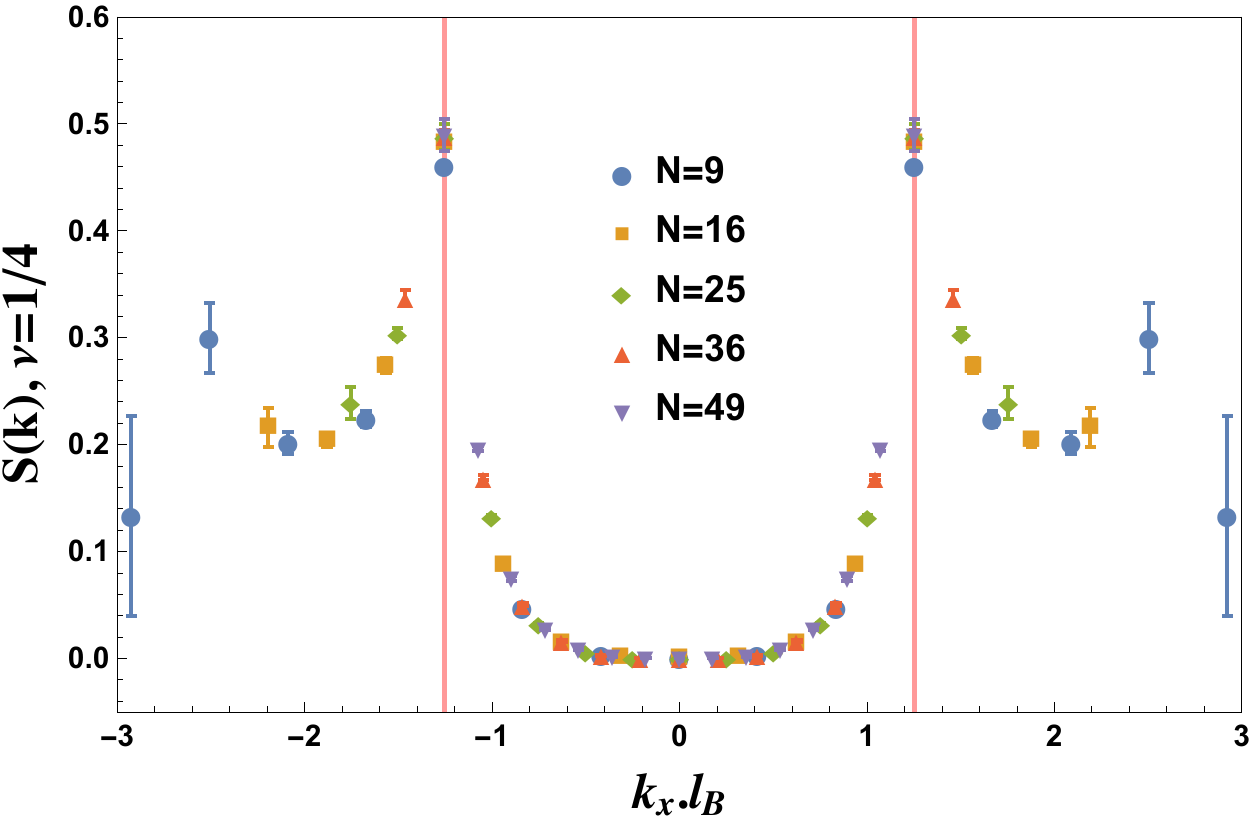}
    \captionsetup{justification=RaggedRight, singlelinecheck=true}
    \caption{Guiding center structure factor $S(k)$ as a function of $k_x$ along $k_y$=0 axes computed from $\Psi^{1/4}_{n=1}$. The plot is obtained after a finite size scaling for MWFs of $\sqrt{N}$$\times$$\sqrt{N}$ square FSs where $N$ is the number of electrons. The red lines are $2k_F.l_B$=$\sqrt{2\pi\nu}$, a value of twice of the Fermi wave vector obtained by applying Luttinger theorem on a square FS. The fact that $S(k)$ plots fit into one curve and the numerical singularities match with the analytical value implies that Luttinger theorem is true for CFLs.}\label{luttinger}
\end{figure}

Some knowledge about the FS can be obtained before an interpretation (whether nonrelativistic HLR fermions or relativistic DFs) of the particles that compose the FS is assigned. The first requirement for being a Fermi liquid is no net magnetic fields on the FS: $\langle B_{\psi}\rangle$$\equiv$$\langle\ee^{ab}\p_aa_b\rangle$=0, where $\langle...\rangle$ denotes mean field expectation value. Second, in Landau Fermi liquid theory, the FS volume is determined by the charge density, known as the Luttinger theorem. It has been conjectured \cite{Son,GeraedtsScience,JainLuttinger} that CFLs satisfy the Luttinger theorem too, {\it i.e.} composite Fermi wave vector is determined by the electrons' filling factor. In Fig.~\ref{luttinger}, we investigated the Luttinger theorem for CFLs by computing the guiding center structure factor $S(k)$ of a \quartnu MWF. $S(k)$ is defined as $\frac{1}{2N_{\phi}}\langle\{\delta\rho(k),\delta\rho(-k)\}\rangle$, where $\delta\rho(k)\equiv\rho(k)-\langle\rho(k)\rangle$ represents the fluctuation of density relative to the ground state mean value and obeys the Girvin-MacDonald-Plazman algebra \cite{gmpl}. As a hallmark of CFL, there are peaks in $S(k)$ and the peak positions are tied to $k$=$2k_F$, twice the Fermi wave vector. The measured $k_F$ agrees with the value predicted from Luttinger theorem, suggesting that the Luttinger theorem applies to CFLs \cite{JainLuttinger,Son,GeraedtsScience}. We will then assume the Luttinger theorem for CFLs and use it as a constraint to derive the effective action.

The PH conjugate of a CFL is supposed to have the same FS size \cite{Maissamph,JainLuttinger,gromovPHsym,levinPHsym}. In the HLR picture, this is because FSs of fCFLs and anti-fCFLs are formed by composite-fermions composite-holes, respectively, whose Fermi level is the same. The DF theory interprets the FS as formed by DFs, which fill DF bands up to the Fermi level determined by the electrons' filling factor through the Luttinger theorem. As a result, the DF density must be
\beqn
\rho_{\psi} = \frac{1}{2m}\frac{1}{2\pi l_B^2},\quad\nu=\begin{cases}1/2m~\text{or}\\1-1/2m.\end{cases}\label{dirac-luttinger}
\eeqn

Taking a variation of $a_0$ and $A_0$ for Eq.~(\ref{undeterminedaction}), the DF density and electron density at mean field level are found to be: $\langle\rho_{\psi}\rangle$=$C_1\frac{B_A}{2\pi}$ and $\langle\rho_A\rangle$=$C_3\frac{B_A}{2\pi}$. Hence, the Luttinger theorem and Hall conductivity determine $C_1$ to be 1/2m and $C_3$ to be $\frac{1}{2}$$-$$\frac{\eta}{2}\frac{m-1}{m}$ at mean field where $\eta$=+1 for fCFLs and =$-$1 for anti-fCFLs. Based on the observation of a $\pi$ peak concentrated at the FS center, we conjecture that the DF is massless. Unlike Son's theory, the presence of a Chern-Simons (CS) term $\frac{C_2}{4\pi}ada$ has a nonuniversal contribution to $\sigma_H$ \cite{KivelsonCFL}, and induces nonzero $U(1)_A$ charge to DFs \cite{ReadLLL,WangSenthilCFL}. The effects of the CS term are canceled provided the FS carries $2\pi C_2$ Berry phase; in other words, the CS term assigns the FS with an extra $2\pi C_2$ Berry phase \cite{Note3}. This $2\pi C_2$ phase, together with the $-\pi$ Berry curvature singularity located at the FS center, comprises the total $-2\pi\nu$ FS berry phase as observed. Based on the fact that Berry phase is odd under PH transformation, we set $C_2$=$\eta\left(\frac{1}{2}-\frac{1}{2m}\right)$. Thus we determined Eq.~(\ref{flux-dirac-action-new}) by the Luttinger theorem, Hall conductivity and FS Berry phase. See Supplementary Material for further discussion.

We will then argue for the Berry curvature distribution presented in Fig.~\ref{berrycurvature} based on a flux-attached DF picture. The CS term induces $U(1)_A$ charge to DFs, making the flux-attached DF theory not in the LLL \cite{ReadLLL,WangSenthilCFL}; the same issue is present in the HLR theory at all fillings \cite{WangSenthilCFL}. In spite of not being a LLL theory, the effect of LLL projection is well known: as a fundamental property of a dipolar electron in a magnetic field, the dipole vector $\bm d$ is perpendicular and the strength proportional to the kinetic momentum vector $\bm k$, i.e. the so-called {\it dipole-momentum locking}; LLL projection, generally speaking, shifts the flux attachment center away from the electron's location to create a dipole. In Son's DF theory, 2 fluxes turn an electron into a DF, and the DF's spin represents a dipole. At $\nu$=1/2$m$, we expect DFs are attached with $(2m$$-$$2)$ residual flux quanta, which after LLL projection are shifted away from the DF's location to form a dipole. To distinguish the $(2m-2)$-flux dipole from the total $2m$-flux dipole (which include spin), we dub the former as residual dipoles. Hence, in flux-attached DF theory, DF has residual dipole-momentum locking in addition to being a spin half spinor.

The residual dipole-momentum locking of the DF has a nontrivial impact on the Berry phase associated with transporting a composite fermion (dipolar DF) in the momentum space ($\bm k$-space). The Berry curvature distribution is predicted to be: (I) $\bm k$-space uniform except at the FS center point $\bm k$=0, (II) where there is an additional $\pi$ Berry phase. The argument goes as follows \footnote{Berry curvature was initially argued by Haldane \cite{MM_Haldane_CFL} to be uniform with Berry curvature density $\eta l_B^{-2}$. Here we found (uniform part) the density is instead $\eta\left(m-1\right)l_B^{-2}$.}.

The dipole-momentum locking provides a nature mapping from the real-space to the $\bm k$-space. The motion of a dipolar DF in $\bm k$-space induces the rotation of the residual dipoles in the real-space \cite{MM_Haldane_CFL}. Since the real-space density is uniform, and since the $\bm k$-space area is proportional to real-space area, (I) is a manifestation of the real-space Aharonov-Bohm effect. The contribution to the FS Berry phase $\Phi_{FS}$ from (I) should be $-(2\pi\nu$$-$$\pi)$ in accordance with the fact that it vanishes at half filling. Then, (II) originates from being massless spin-half DFs. Finite mass tilts the DF's spin away from the 2D plane; hence, mass term $|\mathcal{M}|$ represents how much the FS Berry phase $\Phi_{FS}$ deviates from $-2\pi\nu$. We thus conjecture that the DF mass $|\mathcal{M}|$=0, which we emphasis is not protected by symmetry but instead constrained to take this value by the FS Berry phase Eq.~(\ref{FSBP}).

\begin{figure}[]
    \centering
    \includegraphics[width=0.4\textwidth]{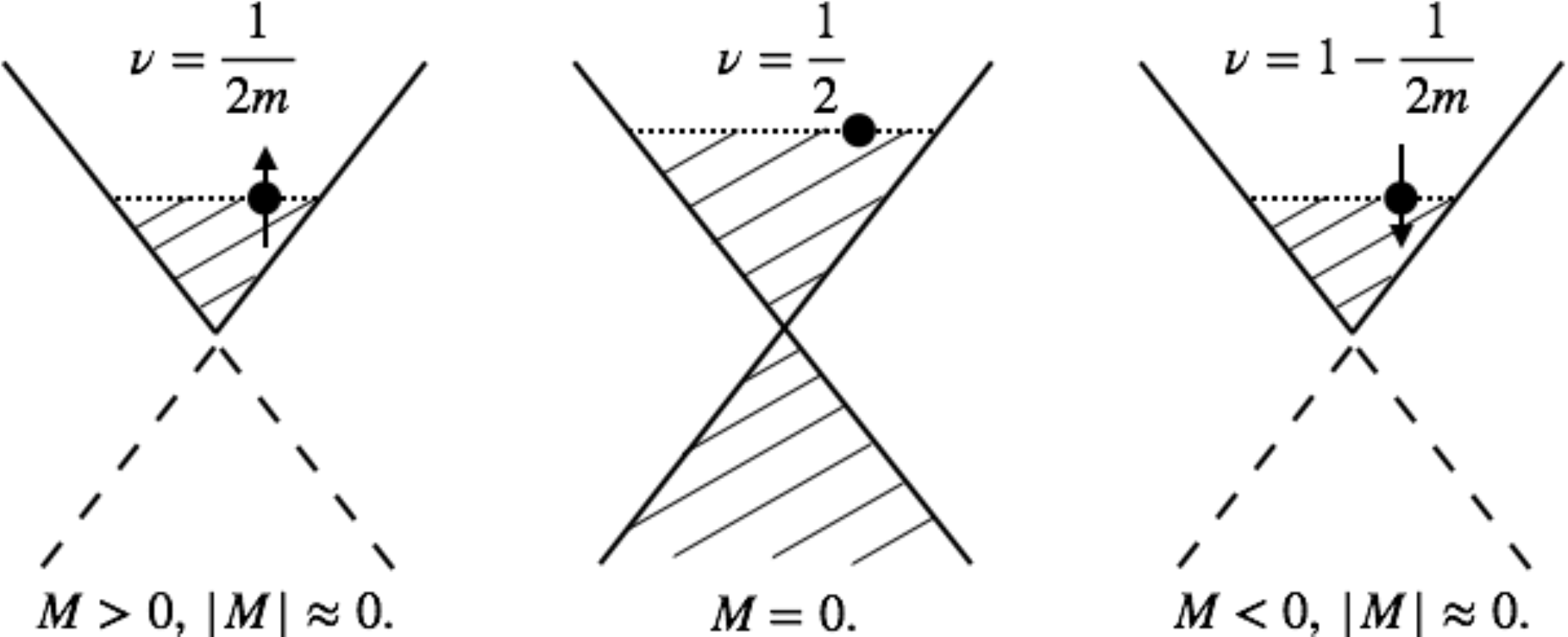}
    \captionsetup{justification=RaggedRight, singlelinecheck=true}
    \caption{Illustrations of FS, particles, band structure in Son's half filled DF theory (middle) and theory $\mathcal{L}_{\psi_+}$ in Eq.~(\ref{flux-dirac-action}). PH acts like time reversal, thus flipping the fluxes (arrows) attached to the DFs (black dots). FS sizes of PH conjugate states are the same, fixed by Luttinger theorem. The valance band, represented as dashed straight line, has been integrated out. The band mass is conjectured to be negligible.
    }\label{lowenergytheory}
\end{figure} 

Finally, we want to make a connection to Son's half filling theory. Because of the lack of PH symmetry when $\nu$$\neq$1/2, Son's theory Eq.~(\ref{DiracFermionAction}) acquires a mass $\mathcal{M}$. To describe low energy excitations around the Fermi surface, the lower band needs to be integrated out, inducing a $\sign{\frac{(\mathcal{M})}{2}}$ level CS term to the action: $\mathcal{L}_{\psi_+}=i\bar\psi'_+\gamma^{\mu}\left(\p_{\mu}-ia'_{\mu}\right)\psi'_+-\frac{1}{2}\frac{1}{2\pi}a'dA+\frac{1}{2}\frac{1}{4\pi}AdA+\frac{\sign{(\mathcal{M}})}{2}\frac{1}{4\pi}a'da'$, where $\psi_+$ describes the upper band fermion only. Fixing the electron density to be $\nu$, the field strength of $a'_{\mu}$ is found to be $\langle B_{a'}\rangle$$\equiv$$\langle\ee^{ab}\p_aa'_b\rangle$=$(1$$-$$2\nu)$$\frac{B_A}{2\pi}$. The upper band density $\langle\rho_{\psi'_+}\rangle$=$\frac{1}{2}\frac{B_A}{2\pi}$$-$$\frac{\sign{(\mathcal{M}})}{2}\frac{\langle B_{a'}\rangle}{2\pi}$ is then $\frac{1}{2m}\frac{B_A}{2\pi}$ for both fCFLs and anti-fCFLs, provided the sign of mass is positive for fCFLs and negative for anti-fCFLs. The filling fraction for $\psi'_{+}$ is then: $\eta(2m-2)^{-1}$ which is again an even denominator fraction. Hence, upon a statistics preserving flux attachment transformation, composite fermions would perceive no magnetic fields on average. After carrying out the flux attachment singular gauge transformation, we arrive at,
\beqn
\mathcal{L}_{\psi_+} &=& i\bar\psi_+\gamma^{\mu}\left(\p_{\mu}-ia_{\mu}\right)\psi_++\frac{\sign(\mathcal{M})}{2m}\frac{1}{4\pi}ada\label{flux-dirac-action}\\
&&-\frac{1}{2m}\frac{1}{2\pi}adA+\left(\frac{1}{2}-\frac{\sign(\mathcal{M})}{2}\frac{m-1}{m}\right)\frac{1}{4\pi}AdA.\nonumber
\eeqn
where $\psi_{+}$ is the upper band flux-attached Dirac fermion field which perceives no net magnetic field at mean field $\langle B_{a}\rangle$$\equiv$$\langle\ee^{ab}\p_aa_b\rangle$=0. The FS of theory $\mathcal{L}_{\psi_+}$ can be viewed as formed by flux-attached DFs, whose density is determined through Luttinger theorem, and whose PH conjugate is attached with same amount but opposite fluxes, see Fig.~\ref{lowenergytheory}. After adding back the valance band to cast the effective action into the standard notion and identifying $\eta$ with $\sign(\mathcal{M})$, we arrive at Eq.~(\ref{flux-dirac-action-new}).

The author is grateful to F. D. M. Haldane for shedding light onto this problem, and acknowledges illuminating discussions with Dam Thanh Son, Todadri Senthil, Hart Goldman, Songyang Pu, Edward Rezayi, Chong Wang, Junyi Zhang and Yahui Zhang. The author appreciates Huan He, Bo Yang, Yi Zhang and Yunqin Zheng for valuable comments on the manuscript. This work was supported by DOE grant No. DE-SC0002140.

\textbf{\emph{Note added.---}}
Recently, Ref.~(\onlinecite{FradkinOneQuarter}) appeared which overlapped with this work and considered the same types of theories from a different but complementary perspective.



\section{Supplementary}
\subsection{Berry Curvature Numerics}
\begin{figure}[H]
    \centering
    \includegraphics[width=0.45\textwidth]{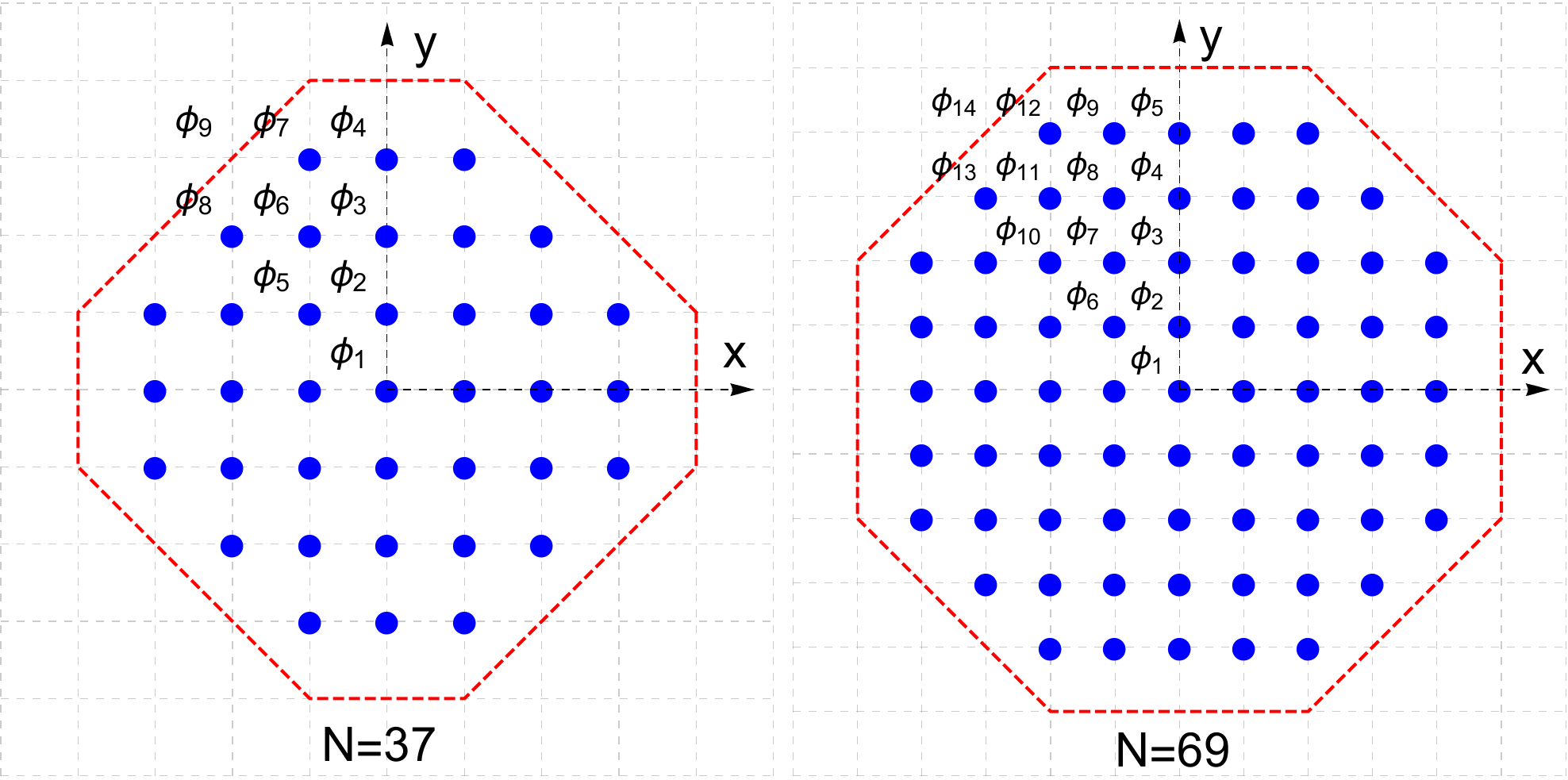}
    \caption{The Fermi sea and discretized Berry curvature $\phi$. The left and right Fermi sea has $N$=37 and $69$ dipoles respectively, on square toruses. The Berry curvatures related by rotation and inversion are not represented. The red dashed line represents the Fermi surface boundary along where transporting a single dipole has $\Phi_{\Gamma}$=$-2\pi\nu$ Berry phase.}\label{fermisea}
\end{figure}

In this section, we provide numerical details about computing Berry curvature, as well as more case studies including N=69 and $\nu$=1/3 states whose underlying particles are bosons. To avoid confusion, we point out that the $\nu=\frac{1}{3}$ states we considered in this work are not gapped Laughlin states, but the bosonic composite Fermi liquid states whose composite fermions are formed by attaching each boson with three flux quanta. The bosonic composite Fermi liquid has been studied in {\it e.g.} Refs.~\cite{pasquirehaldane,ReadLLL}. Fermi sea configurations are shown in Fig.~\ref{fermisea}. The red lines represent the Fermi sea boundary, along which anti-clock-wisely transporting a composite fermion has numerically found to have $-2\pi\nu$ Berry phase. The $\phi$ represents the Berry phase on the corresponding grid, {\it i.e.} the discretized Berry curvature.

Within the composite Fermi liquid (CFL) phase, low energy exact diagonization states can be identified with model wave functions whenever dipoles form clustered configurations. Model wave function with a non-compact Fermi sea is hard to be identified as a single exact diagonization state; instead it might be a linear combination of low energy and excited states. For this reason, we did not consider measuring the Berry phase using a composite hole located deep inside the Fermi sea, nor a composite fermion excited too far away. Instead, we only consider paths $\Gamma$ near the Fermi surface. As described in the main text, the model wave functions we used is as follows,
\beqn
\Psi^{1/2m}_n&=&\det_{ia} M_{ia}\prod_{i<j}^{N}\sigma^{2(m-n)}(z_i-z_j)\nonumber\\
&&\times\prod_k^{2m}\sigma\left(\sum_i^Nz_i-\alpha_k\right)\prod_i^{N}e^{-\frac12z_iz_i^*}.\label{modelwavefunction}
\eeqn
where descriptions about dipoles $\{d\}$, center of mass zeros $\{\alpha\}$ can be found in the main text. In the above model wave function, we only consider model wave functions with $2m$$\geq$$2n$ because otherwise they are $0/0$ indeterminate forms when multiple particles collide onto the same site. The analytical values of these indeterminate forms are well defined, but numerically difficult to evaluate. 

In the main text, we compared the variational energy of model wave function with exact diagonalization out of Coulomb interaction at 1/4 lowest Landau level (LLL). We found good agreement except at $(5, 0)$ sector. Similar phenomena appears at 1/2. At half filling for $10$ particles in the $(5, 0)$ sector, the lowest three Coulomb energies are $-0.4649$, $-0.4634$, $-0.4567$ [in units of $e^2/\ee l_B$]. The lowest variational energy of the model wave function, which corresponds to the most compact Fermi sea consisting of $10$ dipoles on a square torus, is $-0.4625\pm0.0006$. Same as \quartnu, this variational energy is close to the next lowest rather than the lowest energy. We hence postulate that it is due to the difficulty of packing 10 dipoles into a compact Fermi sea in the $(5, 0)$ sector on a square torus. 

The determinant with Jastrow factors of power two vanishes identically when $\bar d$ approaches Fermi sea center, if $N$ is even and $\{d\}$ is inversion symmetric \cite{Jie_MonteCarlo}. This gives some hints that two fluxes attachment turns an electron into a Dirac fermion. Our intuition for the Berry curvature distribution from a model wave function point of view is as follows: the determinant with 2 Jastrow factors corresponds to the $\pi$ singularity at Fermi sea center, while the $-(2\pi/2m-\pi)$ Berry phase corresponds to the 2m$-$2 power of Jastrow factors. From the point of view of Fermi liquid, the former is attributed to the gapless Dirac node or the spin-orbit locking on the Fermi surface as a Fermi surface property, while the latter is attributed to the Berry phase distributed over the Fermi sea which is absent in the Landau Fermi liquid description. Therefore we call the $-(2\pi/2m-\pi)$ Berry phase as an \emph{extra prescription}. From effective theory point of view, this extra prescribed Berry phase is attributed to the Chern-Simons term, which has a non-universal contribution to the Hall conductivity and which we want to cancel out by the Fermi sea Berry phase according to the theory of anomalous Hall effects in two dimension. See the last section in this supplementary material for more details about this point. From the dipole picture, 2 fluxes turn an electron into a Dirac fermion, while the rest $(2m-2)$ fluxes are attached to Dirac fermions, which after LLL projection form residual dipole-momentum lockings. The residual dipole-momentum locking is supposed to give a uniform Berry curvature density over the Fermi sea, as argued in the main text.

In the numerical Berry phase part, we adopt the lattice Monte Carlo \cite{Jie_MonteCarlo,haldanemodularinv,haldaneholomorphic} to extract the many body Berry phase $\tilde\Phi_{\Gamma}$, which is defined as \cite{scottjiehaldane},
\beqn
|D|~e^{i\tilde\Phi_{\Gamma}} &=& \Tr\prod_{i\in\Gamma}\langle\Psi(\bm K'_i)|\rho(\bm K'_i-\bm K_i)|\Psi(\bm K_i)\rangle,\label{manybodybp}
\eeqn
where $|D|$ is a real number representing the many body amplitude. $|\Psi(\bm K)\rangle$ is a many body model wave function, Eq.~(\ref{modelwavefunction}). The $\bm K$ is the many body momentum which is essentially the sum of all single dipole momentums. The $\rho(\bm q)$=$\sum_i^Ne^{i\bm q\times\bm R_i}$ is the density operator projected into the LLL and satisfies the Girvin-MacDonald-Plazman algebra \cite{gmpl}. The $\bm R_i$ are projected electron coordinates, {\it i.e.} guiding centers. With $i,j$ labeling electrons, $a,b$ labeling spacial directions, $l_B$=$\sqrt{\hbar/|eB|}$ being the magnetic length, and $\epsilon^{xy}$=$-\epsilon^{yx}$=1 representing the anti-symmetric symbol, the guiding center coordinates satisfy a non-commutative relation as $[R_i^a, R_j^b]$=$-i\ee^{ab}l_B^2$. 

We found the many body Berry phase, besides the genuine phase $\Phi_{\Gamma}$, has a path dependent piece $(i)^{N_+-N_-}$ where $N_{+}$/ $N_{-}$ is the anti-clock-wised/ clock-wised step number in $\Gamma$ relative to the Fermi sea center:
\beqn
e^{i\tilde\Phi_{\Gamma}} &=& (i)^{N_+-N_-}\cdot e^{i\Phi_{\Gamma}}.\label{pathdependence}
\eeqn

At half filling \cite{scottjiehaldane}, this phase factor has been argued based on particle hole and reflection symmetry. In the next section, we justify such phase factor for generic filling fractions.

\begin{figure}[]
    \centering
    \hspace*{-0.025\textwidth}\begin{subfigure}{0.45\textwidth}
    \centering
    \includegraphics[width=0.75\textwidth]{mapoutdata4_37_new.pdf}
    \end{subfigure}
    \hspace*{-0.025\textwidth}\begin{subfigure}{0.45\textwidth}
    \centering
    \includegraphics[width=0.75\textwidth]{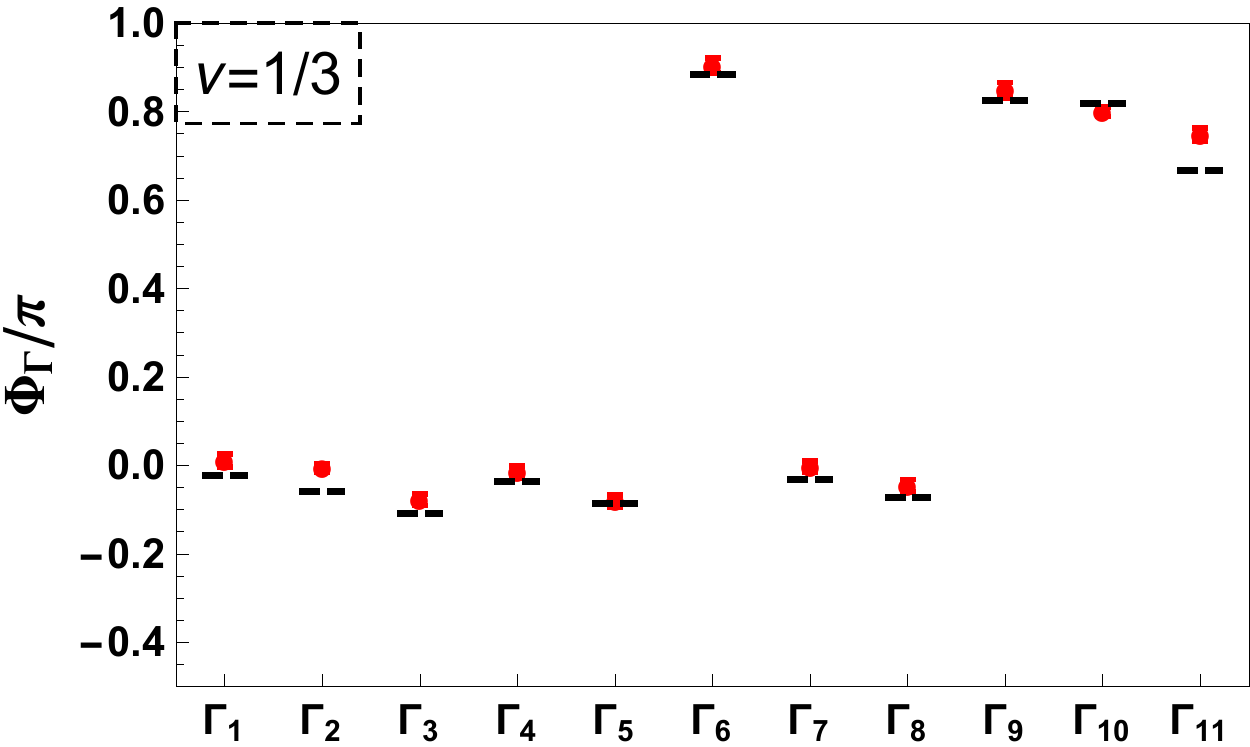}
    \end{subfigure}
    \captionsetup{justification=RaggedRight, singlelinecheck=true}
    \caption{The Berry phases associated with paths $\Gamma_1, ...$ listed in Table~\ref{mapoutfstab} for $N$=37 Fermi sea computed from $\nu$=1/4 [upper panel] and $\nu$=1/3 [lower panel] CFL model wave functions.}\label{mapoutbp37}
\end{figure}
\begin{figure}[]
    \centering
    \hspace*{-0.025\textwidth}\begin{subfigure}{0.45\textwidth}
    \centering
    \includegraphics[width=0.75\textwidth]{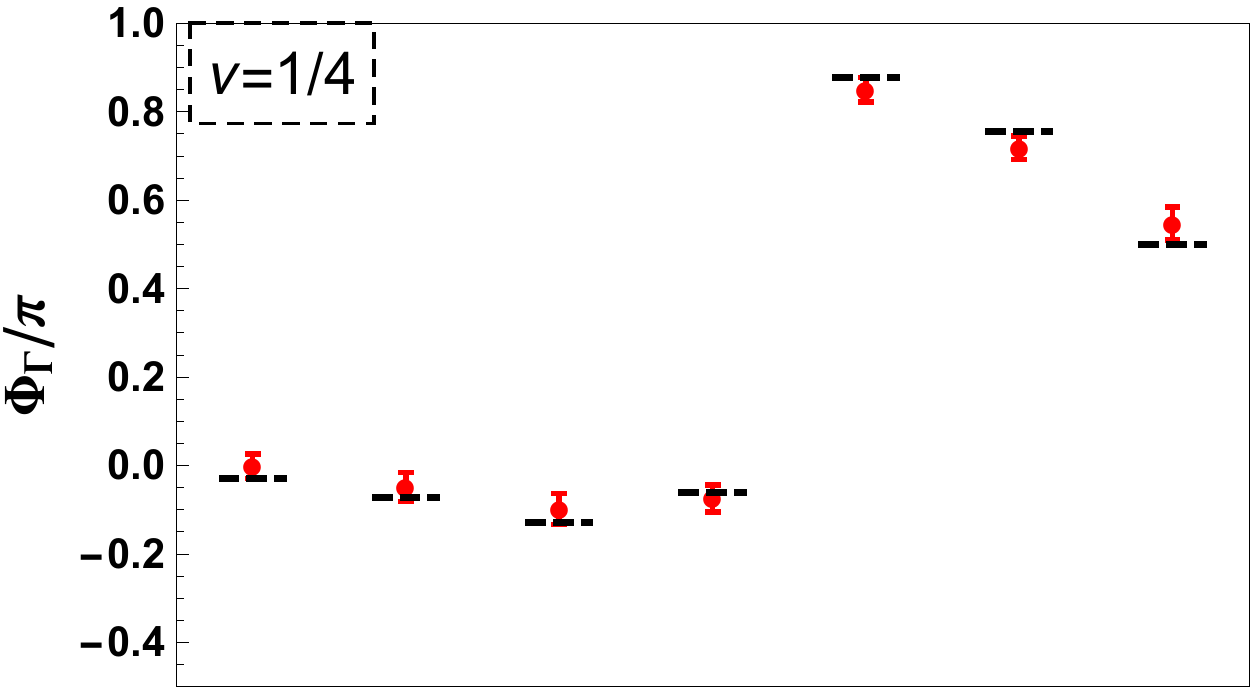}
    \end{subfigure}
    \hspace*{-0.025\textwidth}\begin{subfigure}{0.45\textwidth}
    \centering
    \includegraphics[width=0.75\textwidth]{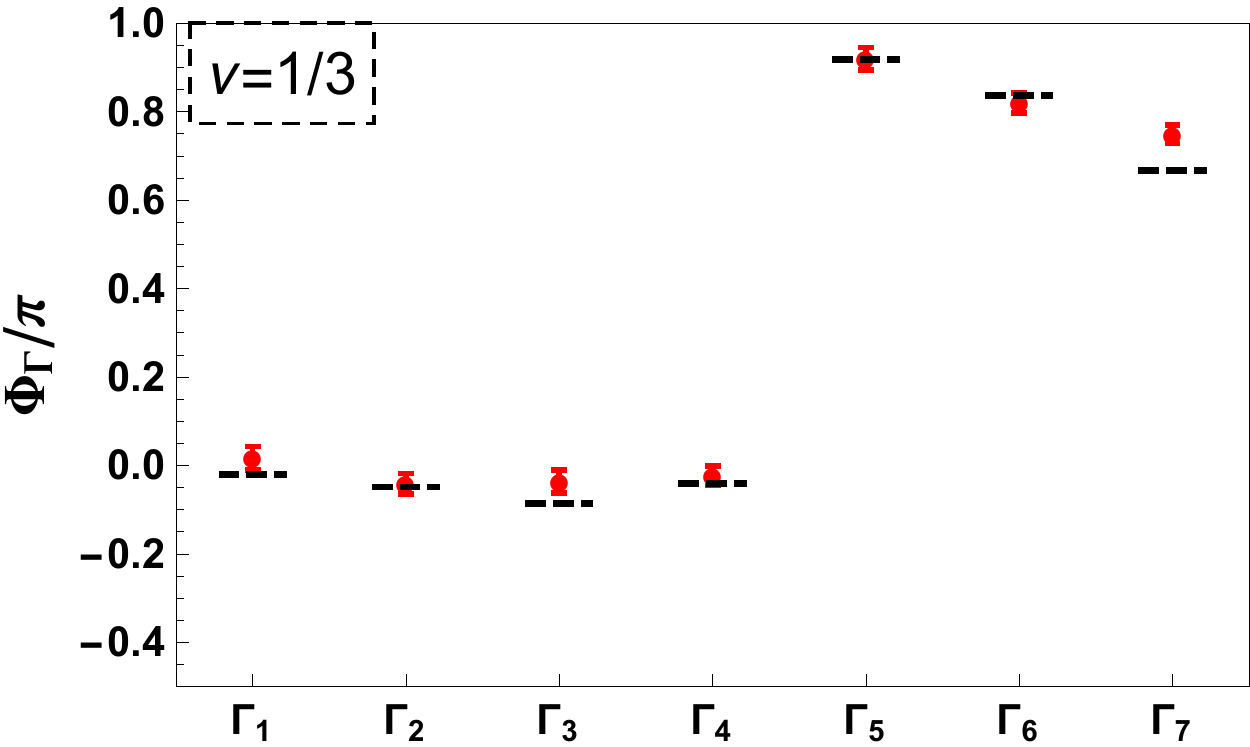}
    \end{subfigure}
    \captionsetup{justification=RaggedRight, singlelinecheck=true}
    \caption{Same as Fig.~\ref{mapoutbp37}, but for $N$=69 Fermi sea. In Fig.~\ref{mapoutbp37} and Fig.~\ref{mapoutbp69}, black dashed lines are Berry phases according to Eq.~(\ref{curvatureonequarter_appendix}) and the red dots stand for Monte Carlo values which can be found in Table~\ref{mapoutfstab}.}\label{mapoutbp69}
\end{figure}

The physical Berry phase $\Phi_{\Gamma}$ is an area weighted sum of the discretized curvature $\phi$. The numerical values of $\Phi_{\Gamma}$ computed on the corresponding paths $\Gamma$ from $\nu$=1/4 and $1/3$ model wave functions are listed in TABLE~\ref{mapoutfstab}. We found an empirical formula for the Berry phase, which is the key result in the numerical part of this work, as follows [transport composite fermion clock-wisely],
\beqn
\Phi_{\Gamma} = \delta_{\Gamma}\cdot\pi + (2\pi\nu-\pi)\cdot A_{\Gamma}/A_{FS}.\label{curvatureonequarter_appendix}
\eeqn
where $\delta_{\Gamma}$ is the winding number of the path $\Gamma$ relative to the Fermi sea center: it is $+1$ if $\Gamma$ encloses Fermi sea center once and $0$ if not. $A_{\Gamma}$ and $A_{FS}$ are the momentum space area enclosed by path $\Gamma$ and Fermi sea area respectively. The importance of Eq.~(\ref{curvatureonequarter_appendix}) has already been emphasized in the main text: it implies a uniform Berry curvature together with a $\pi$ singularity at Fermi sea center. Motivated by Eq.~(\ref{curvatureonequarter_appendix}), a flux-attached Dirac fermion theory is proposed in the main text. In Fig.~\ref{mapoutbp37} and Fig.~\ref{mapoutbp69}, we compare the $\Phi_{\Gamma}$ value computed from numerics and the analytical value calculated from Eq.~(\ref{curvatureonequarter_appendix}) for various paths on the $N$=37 and 69 Fermi seas. The excellent agreement indicates that Eq.~(\ref{curvatureonequarter_appendix}) makes sense.

To better visualize the Berry curvatures, we did a linear regression and mapped out Berry curvatures on $N$=37 Fermi sea. The linear relations of $\Phi_{\Gamma}$ and $\phi$ can be found in the following Eq.~(\ref{X37}). The Berry curvature plot is shown in the main text.

{\small
\beqn
\Phi_{\Gamma{1}} &=& 2\phi_4+\phi_7 \label{X37}\\
\Phi_{\Gamma{2}} &=& 2\phi_3+2\phi_4+2\phi_6+\phi_7+\phi_8 \nonumber\\
\Phi_{\Gamma{3}} &=& 2\phi_2+2\phi_3+2\phi_4+2\phi_5+4\phi_6+2\phi_7+\phi_8 \nonumber\\
\Phi_{\Gamma{4}} &=& 2\phi_3+2\phi_6+\phi_8\nonumber\\
\Phi_{\Gamma{5}} &=& 2\phi_2+2\phi_3+2\phi_5+4\phi_6+\phi_7+\phi_8 \nonumber\\
\Phi_{\Gamma{6}} &=& 4\phi_1+4\phi_2+4\phi_3+4\phi_4 \nonumber\\
\Phi_{\Gamma{7}} &=& \phi_3+2\phi_4+\phi_6+\phi_5/2 \nonumber\\
\Phi_{\Gamma{8}} &=& \phi_2+3\phi_3+4\phi_4+\phi_5+\phi_6 \nonumber\\
\Phi_{\Gamma{9}} &=& 4\phi_1+6\phi_2+2\phi_3+2\phi_5+4\phi_6+2\phi_7+2\phi_8+2\phi_9\nonumber\\
\Phi_{\Gamma{10}}&=& 4\phi_1+6\phi_2+89\phi_3/15+11\phi_4/3+2\phi_5+41\phi_6/15 \nonumber\\
&+& 3\phi_7/5+\phi_8/15. \nonumber\\
\Phi_{\Gamma{11}}&=& 4\phi_1+8\phi_2+8\phi_3+8\phi_4+4\phi_5+8\phi_6 \nonumber\\
&+& 4\phi_7+2\phi_8. \nonumber
\eeqn
}

\begin{table}[]
\begin{center}
\begin{tabular}{ |>{\scriptsize}c|>{\scriptsize}c|*{4}{>{\scriptsize}c|} }
\hline
\multicolumn{2}{|c|}{$N=37$, path $\Gamma$ ($\bm K_i'\rightarrow\bm K_i$)} & $\Phi^{\nu=\frac{1}{3}}_{\Gamma, n=1}/\pi$ & $\Phi^{\nu=\frac{1}{4}}_{\Gamma, n=1}/\pi$ \\ \hline
$\Gamma_1$ & \shortstack{$\begin{aligned}(1,4) &\rightarrow (-1,4) \rightarrow (-2,3) \\[-5pt] &\rightarrow (2,3) \rightarrow (1,4)\end{aligned}$} & $0.011\pm0.015$ & $-0.004\pm0.021$ \\ \hline
$\Gamma_2$ & \shortstack{$\begin{aligned}(1,4) &\rightarrow (-1,4) \rightarrow (-3,2) \\[-5pt] &\rightarrow (3,2) \rightarrow (1,4)\end{aligned}$} & $-0.006\pm0.010$ & $-0.067\pm0.023$ \\ \hline
$\Gamma_3$ & \shortstack{$\begin{aligned}(1,4) &\rightarrow (-1,4) \rightarrow (-4,1) \\[-5pt] &\rightarrow (4,1) \rightarrow (1,4)\end{aligned}$} & $-0.077\pm0.013$ & $-0.150\pm0.018$ \\ \hline
$\Gamma_4$ & \shortstack{$\begin{aligned}(2,3) &\rightarrow (-2,3) \rightarrow (-3,2) \\[-5pt] &\rightarrow (3,2) \rightarrow (2,3)\end{aligned}$} & $-0.015\pm0.014$ & $-0.056\pm0.023$ \\ \hline
$\Gamma_5$ & \shortstack{$\begin{aligned}(2,3) &\rightarrow (-2,3) \rightarrow (-4,1) \\[-5pt] &\rightarrow (4,1) \rightarrow (2,3)\end{aligned}$} & $-0.081\pm0.014$ & $-0.152\pm0.022$ \\ \hline
$\Gamma_6$ & \shortstack{$\begin{aligned}(4,1) &\rightarrow (-4,1) \rightarrow (-4,-1) \\[-5pt] &\rightarrow (4,-1) \rightarrow (4,1)\end{aligned}$} & $0.905\pm0.017$ & $0.832\pm0.018$ \\ \hline
$\Gamma_7$ & \shortstack{$\begin{aligned}(0,4) &\rightarrow (-4,0) \rightarrow (-4,-1) \\[-5pt] &\rightarrow (1,4) \rightarrow (0,4)\end{aligned}$} & $-0.002\pm0.013$ & $-0.029\pm0.019$ \\ \hline
$\Gamma_8$ & \shortstack{$\begin{aligned}(0,4) &\rightarrow (-4,0) \rightarrow (-4,-2) \\[-5pt] &\rightarrow (2,4) \rightarrow (0,4)\end{aligned}$} & $-0.046\pm0.014$ & $-0.098\pm0.021$ \\ \hline
$\Gamma_9$ & \shortstack{$\begin{aligned}(2,4) &\rightarrow (-4,-2) \rightarrow (-2,-4) \\[-5pt] &\rightarrow (4,2) \rightarrow (2,4)\end{aligned}$} & $0.848\pm0.018$ & $0.741\pm0.021$ \\ \hline
$\Gamma_{10}$ & \shortstack{$\begin{aligned}(1,4) &\rightarrow (-1,4) \rightarrow (-4,1) \\[-5pt] &\rightarrow (4,1) \rightarrow (1,4)\end{aligned}$} & $0.800\pm0.011$ & $0.688\pm0.019$ \\ \hline
$\Gamma_{11}$ & \shortstack{$\begin{aligned}(1,4) &\rightarrow (-1,4) \rightarrow (-4,1) \\[-5pt] &\rightarrow (-4,-1) \rightarrow (-1,-4) \\[-5pt] &\rightarrow (1,-4) \rightarrow (4,-1) \\[-5pt] &\rightarrow (4,1) \rightarrow (1,4)\end{aligned}$} & $0.748\pm0.015$ & $0.535\pm0.017$ \\ \hline
\hline
\multicolumn{2}{|c|}{$N=69$, path $\Gamma$ ($\bm K_i'\rightarrow\bm K_i$)} & $\Phi^{\nu=\frac{1}{3}}_{\Gamma, n=1}/\pi$ & $\Phi^{\nu=\frac{1}{4}}_{\Gamma, n=1}/\pi$ \\ \hline
$\Gamma_1$ & \shortstack{$\begin{aligned}(2,5) &\rightarrow (-2,5) \rightarrow (-3,4) \\[-5pt] &\rightarrow (3,4) \rightarrow (2,5)\end{aligned}$} & $0.016\pm0.026$ & $-0.002\pm0.028$ \\ \hline
$\Gamma_2$ & \shortstack{$\begin{aligned}(2,5) &\rightarrow (-2,5) \rightarrow (-4,3) \\[-5pt] &\rightarrow (4,3) \rightarrow (2,5)\end{aligned}$} & $-0.042\pm0.024$ & $-0.049\pm0.033$ \\ \hline
$\Gamma_3$ & \shortstack{$\begin{aligned}(2,5) &\rightarrow (-2,5) \rightarrow (-5,2) \\[-5pt] &\rightarrow (5,2) \rightarrow (2,5)\end{aligned}$} & $-0.036\pm0.027$ & $-0.098\pm0.035$ \\ \hline
$\Gamma_4$ & \shortstack{$\begin{aligned}(5,2) &\rightarrow (-5,2) \rightarrow (-5,1) \\[-5pt] &\rightarrow (5,1) \rightarrow (5,2)\end{aligned}$} & $-0.024\pm0.022$ & $-0.074\pm0.031$ \\ \hline
$\Gamma_5$ & \shortstack{$\begin{aligned}(5,1) &\rightarrow (-5,1) \rightarrow (-5,-1) \\[-5pt] &\rightarrow (5,-1) \rightarrow (5,1)\end{aligned}$} & $0.919\pm0.026$ & $0.849\pm0.028$ \\ \hline
$\Gamma_6$ & \shortstack{$\begin{aligned}(5,2) &\rightarrow (-5,2) \rightarrow (-5,-2) \\[-5pt] &\rightarrow (5,-2) \rightarrow (5,2)\end{aligned}$} & $0.820\pm0.022$ & $0.719\pm0.026$ \\ \hline
$\Gamma_{7}$ & \shortstack{$\begin{aligned}(2,5) &\rightarrow (-2,5) \rightarrow (-5,2) \\[-5pt] &\rightarrow (-5,-2) \rightarrow (-2,-5) \\[-5pt] &\rightarrow (2,-5) \rightarrow (5,-2) \\[-5pt] &\rightarrow (5,2) \rightarrow (2,5)\end{aligned}$} & $0.749\pm0.021$ & $0.547\pm0.038$ \\ \hline
\end{tabular}
\captionsetup{justification=RaggedRight, singlelinecheck=true}
\caption{
Paths $\Gamma$ and phases $\Phi_{\Gamma}$ for N=37 and $69$ Fermi sea. See Eq.~(\ref{manybodybp}) for definitions of $\bm K$ and $\bm K'$. For example, in $\Gamma_1$ of N=37, the Berry phase is computed as follows: $|D|e^{i\tilde\Phi_{\Gamma}}=\Tr\langle\Psi_{(1,4)}|\rho|\Psi_{(-1,4)}\rangle$ $\langle\Psi_{(-1,4)}|\rho|\Psi_{(-2,3)}\rangle$ ... $\langle\Psi_{(2,3)}|\rho|\Psi_{(1,4)}\rangle$, $\Phi_{\Gamma}=\tilde\Phi_{\Gamma}-(3-1)\cdot\pi/2$. Note that the extra dipole is in fact transported clock-wisely.
 }\label{mapoutfstab}
\end{center}
\end{table}

\subsection{Path-dependent Phase Factor at Generic Filling Factor}
In this section, we show that at any filling fraction, the path dependent phase factor in Eqn.~(\ref{pathdependence}) is necessary to make the many body Berry phase transform consistently under particle hole conjugation and path orientation flipping. We show that for the physical Berry phase to be odd under particle hole conjugation, the many body Berry phase has to have an unphysical path dependent phase. We assume the path dependent phase depends on the details of paths only through anti-clock-wised/ clock-wised path numbers $N_{\pm}$, and then show it has to be the form as shown in Eqn.~(\ref{pathdependence}).

We denote $\mathcal{A}$ as the particle hole conjugation operator, which is anti-unitary. Let $|\Phi\rangle$ and $|\Psi\rangle$ be two many body states. The algebra of anti-unitary operators [see Chap. \text{XV}, \emph{Quantum Mechanics} by A. Messiah] states that,
\beqn
\langle\Phi|(\mathcal{A}|\Psi\rangle) &=& [(\langle\Phi|\mathcal{A})|\Psi\rangle]^*.\label{antiu1}\\
\langle\mathcal{A}\Psi| &\equiv& (|\mathcal{A}\Psi\rangle)^{\dag} = (\langle\Psi|\mathcal{A}^{\dag}).\label{antiu2}
\eeqn
where $(...)$ explicitly indicates the operator acts to the left or to the right. Consider $\langle\Psi_1|\rho(\bm q_{12})|\Psi_2\rangle$ [we wrote $\langle\Psi_1|\rho|\Psi_2\rangle$ for short] where $\bm q_{12}$ matches the momentum difference of $|\Psi_{1,2}\rangle$. The $\mathcal{A}^{\dag}\mathcal{A}$ is an identity operator when acting to the right of a state. By operator insertion, we have,
\beqn
\langle\Psi_1|\rho|\Psi_2\rangle &=& \langle\mathcal{A}^{\dag}\mathcal{A}\Psi_1|\rho|\mathcal{A}^{\dag}\mathcal{A}\Psi_2\rangle\nonumber\\
&=& (\langle\mathcal{A}\Psi_1|\mathcal{A})|\rho\mathcal{A}^{\dag}\mathcal{A}\Psi_2\rangle\nonumber\\
&=& [\langle\mathcal{A}\Psi_1|\mathcal{A}\rho\mathcal{A}^{\dag}\mathcal{A}\Psi_2\rangle]^*.\label{dev1}
\eeqn
where we used Eqn.~(\ref{antiu1}) and Eqn.~(\ref{antiu2}) in the second and third line respectively. Under particle hole conjugation, the density operator transforms as $\mathcal{A}\rho(\bm q)\mathcal{A}^{\dag}$ = $-\rho(-\bm q)$. From Eqn.~(\ref{dev1}) we get,
\beqn
\langle\Psi_1|\rho|\Psi_2\rangle = (-1)\left(\langle\mathcal{A}\Psi_1|\rho|\mathcal{A}\Psi_2\rangle\right)^*.\label{dev2}
\eeqn

Keep in mind that although we omitted the $\bm q$ dependence in $\rho(\bm q)$, the $\rho$ on the left and right hand side above have opposite momentum, since particle hole conjugation maps many body momentum $\bm K$ to $-\bm K$. Eqn.~(\ref{dev2}) implies the many body Berry phase (which is a product of $\langle\Psi_1|\rho|\Psi_2\rangle$ along a path $\Gamma$) must satisfy:
\beqn
e^{i\tilde\Phi_{\Gamma}} = (-1)^N~\left(e^{i\tilde\Phi^{\mathcal{A}}_{\Gamma_{\mathcal{A}}}}\right)^*.\label{constraint}
\eeqn
where $N$ is the total path number, {\it i.e.} number of density operators. The $\tilde\Phi^{\mathcal{A}}_{\Gamma}$ denotes the many body Berry phase for the particle hole partner on path $\Gamma$. The path with subscription $\Gamma_{\mathcal{A}}$ is obtained from $\Gamma$ by inversion hence has the same orientation as $\Gamma$. The Eqn.~(\ref{constraint}) indicates each density operator inserts a pure imaginary phase \cite{scottjiehaldane}. The $(-1)^N$ factor implies that many body Berry phase factor $e^{i\tilde\Phi_{\Gamma}}$ must contain a path dependent phase. We hence write $e^{i\tilde\Phi_{\Gamma}}$ as $e^{i\Phi_{N_+, N_-}}~e^{i\Phi_{\Gamma}}$ where $e^{i\Phi_{\Gamma}}$ is the physical path-independent Berry phase, and $e^{i\Phi_{N_+,N_-}}$ is the unphysical path-dependent part which we \emph{ad hocly assume} dependents on the path only through the number of anti-clock-wised/ clock-wised path numbers $N_{\pm}$. Plus this into Eqn.~(\ref{constraint}), we have:
\beqn
e^{i\Phi_{N_+,N_-}}e^{i\Phi_{\Gamma}} = (-1)^N~e^{-i\Phi_{N_+,N_-}}e^{-i\Phi^{\mathcal{A}}_{\Gamma_{\mathcal{A}}}}.
\eeqn

The physical Berry phase is odd under particle hole conjugation (this is true based on the knowledge of Bloch states and time reversal): $\Phi^{\mathcal{A}}_{\Gamma_{\mathcal{A}}}$ = $-\Phi_{\Gamma}$. So the path dependent phase factor must satisfy:
\beqn
e^{i\Phi_{N_+,N_-}} = (-1)^N~e^{-i\Phi_{N_+,N_-}}.\label{dev3}
\eeqn

Next we consider two path $\Gamma$ and $\Gamma'$ whose $\bm K$ points are the same but with opposite orientation, {\it i.e.} opposite arrow linking adjacent $\bm K$ points. Thus $N_{\pm}$ of $\Gamma$ becomes $N_{\mp}$ of $\Gamma'$. We denote the many body Berry phase of $\Gamma'$ as $\prod\langle\Psi_2|\rho|\Psi_1\rangle$, which is just the complex conjugate of that for $\Gamma$: $\left(\prod\langle\Psi_1|\rho|\Psi_2\rangle\right)^*$ = $\prod\langle\Psi_2|\rho|\Psi_1\rangle$. Therefore, we obtained,
\beqn
\left(e^{i\Phi_{N_+,N_-}}~e^{i\Phi_{\Gamma}}\right)^* &=& e^{i\Phi_{N_-,N_+}}~e^{i\Phi_{\Gamma'}},\nonumber\\
&=& e^{i\Phi_{N_-,N_+}}~e^{-i\Phi_{\Gamma}}.\label{dev4}
\eeqn
where we used the fact that $\Phi_{\Gamma}$ is odd under flipping the path orientation. The above equation implies that the phase factor is odd under swapping $N_{\pm}$:
\beqn
e^{i\Phi_{N_+,N_-}} = e^{-i\Phi_{N_-,N_+}}.\label{dev5}
\eeqn

Eqn.~(\ref{dev3}) and Eqn.~(\ref{dev5}) are the consistency conditions we derived for the path dependent phase factor. They imply that it can either be $(i)^{N_+-N_-}$ or $(-i)^{N_+-N_-}$.

We are unable to distinguish $\pm i$, but we conjecture it is determined by the sign of the magnetic field, {\it i.e.} consistency rule under time reversal conjugation. It might be derived through the GMP algebra $[\rho(\bm q_1), \rho(\bm q_2)]$ = $-2i\sign(B)\sin\left(\frac{\bm q_1\times\bm q_2}{2l_B^2}\right)\rho(\bm q_1+\bm q_2)$, from which we intuitively see that each density operator corresponds to a factor of $(-i\sign(B))$. In this work, we have chosen the convention $B<0$ so that the phase factor is $(i)^{N_+-N_-}$ which is consistent with the numerics. We leave a more rigorous proof for this path dependent phase factor, as well as the proof for the empirical fact that matrix elements $\langle\Psi_1|\rho|\Psi_2\rangle$ vanish identically for paths normal to the Fermi surface, as future work.

\subsection{Composite Fermion Hall Conductivity}
In this section, we consider effective theories of the following type Eq.~(\ref{RPA1}) as a more detailed explanation for the paragraph following Eq.~(7) of the main text. Discussion here essentially follows \cite{KivelsonCFL,haldaneanomaloushall,Son}.
\beqn
\mathcal{L} &=& \mathcal{L}_0(\psi, a) + \frac{1}{2m}\frac{1}{4\pi}(A-a)d(A-a).\label{RPA1}
\eeqn
where $\mathcal{L}_0(\psi, a)$ is the composite fermion action [with interaction terms included]. It can be either non-relativistic or relativistic. The electron and composite-fermion response functions are denoted as $\Pi^{\mu\nu}$ and $\tilde\Pi^{\mu\nu}$ respectively.
\beqn
\mathcal{L}(\psi,a,A) &\equiv& \int\frac{d^3q}{(2\pi)^3}~\frac{1}{2}A_{\mu}^*(q)\Pi^{\mu\nu}(q)A_{\nu}(q).\\
\mathcal{L}_0(\psi, a) &\equiv& \int\frac{d^3q}{(2\pi)^3}~\frac{1}{2}a_{\mu}^*(q)\tilde\Pi^{\mu\nu}(q)a_{\nu}(q).
\eeqn
We will find out how $\Pi^{\mu\nu}$ depends on $\tilde\Pi^{\mu\nu}$. In the following, we work on flat spacetime: metric $g_{ab}$ is constant. We choose Coulomb gauge $A_0$=0, $g^{ab}q_aA_b(x)$=0, hence $\Pi^{\mu\nu}$ has spacial components only [same for $a_{\mu}(x)$ and $\tilde\Pi$].

The $\Pi^{ab}$ has the following decomposition into longitudinal part $\Pi_T$, transversal part $\Pi_T$ and Hall part $\Pi_H$ as follows \cite{Son},
\beqn
\Pi^{ab}(q) = \frac{q^aq^b}{q^2}\Pi_L + \left(g^{ab}-\frac{q^aq^b}{q^2}\right)\Pi_T + \ee^{ab}\Pi_H.\label{PIform}
\eeqn
where $q^a$$\equiv$$g^{ab}q_b$, $q^2$$\equiv$$q_aq^a$. Its inverse is,
\beqn
\det\Pi\cdot\Pi^{-1}_{ab} &=& \frac{q_aq_b}{q^2}\Pi_T + \left(g_{ab}-\frac{q_aq_b}{q^2}\right)\Pi_L - \ee^{ab}\Pi_H.\nonumber
\eeqn
where $\det\Pi$ is the determinant $\det\Pi=\Pi_T\Pi_L - \Pi_H^2$. Define Chern-Simons response $\ee^{ab}(q)$, and full response $\bar\Pi^{ab}(q)$ as,
\beqn
\ee^{ab}(q) &\equiv& -\frac{i\omega}{2\pi}\ee^{ab},\nonumber\\
\bar\Pi^{ab}(q) &\equiv& \tilde\Pi^{ab}(q) + \frac{1}{2m}\ee^{ab}(q).
\eeqn
Integrating out $a_{a}(p)$ and $a_{a}^*(p)$ in Eq.~(\ref{RPA1}), we get,
\beqn
\Pi^{ab}(q) &=& -\left(\frac{1}{2m}\right)^2\ee^{aa'}(q)~\bar\Pi^{-1}_{a'b'}(q)~\ee^{b'b}(q) + \frac{1}{2m}\ee^{ab}(q).\nonumber
\eeqn
With the help of Eq.~(\ref{PIform}), $\Pi^{ab}$ can be expressed as,
\beqn
\Pi^{ab}(q) = -\left(\frac{1}{2m}\right)^2\frac{\omega^2}{(2\pi)^2}\frac{\bar\Pi^T(q)}{\det\bar\Pi} + \frac{1}{2m}\ee^{ab}(q).
\eeqn
with which the electron Hall conductivity is derived,
\beqn
2\pi\sigma_H(q) &\equiv& -\frac{\pi}{i\omega}\ee_{ab}\Pi^{ab}(q),\\
&=& \left(\frac{1}{2m}\right)^2\left(\frac{\omega}{2\pi}\right)^2\frac{2\pi\sigma^{CF}_H(q) + \frac{1}{2m}}{\det\bar\Pi} + \frac{1}{2m}.\nonumber
\eeqn
Hence Hall conductivity gets a non-universal correction from composite fermions when integrating out internal gauge fields, except when $2\pi\sigma_H^{CF}$=$-$1/2$m$.

\subsubsection{Son's action}
In Son's Dirac fermion theory, $m$=1 and,
\beqn
\mathcal{L}_0 = i\bar\psi\gamma^{\mu}(\p_{\mu}-ia_{\mu})\psi - \frac{1}{2}\frac{1}{4\pi}ada + ...
\eeqn
where $...$ represents interaction terms. A single massless Dirac fermion has parity anomaly: it cannot be quantized in a gauge invariant way, unless a half level Chern-Simons term is included which breaks time reversal symmetry explicitly \cite{dualityweb}. Such half Hall conductivity is equivalently attributed to the $\pi$ Berry curvature singularity of Dirac fermion, which we emphasis should be regarded as a singularity in Berry curvature rather than a Berry phase [since Berry phase is defined with a gap]. To conclude, due to parity anomaly or $\pi$ Berry curvature singularity, the Dirac composite fermion has the required $2\pi\sigma_H^{CF}$=$-$1/2.

\subsubsection{HLR action}
In HLR theory,
\beqn
\mathcal{L}_0 = i\psi^{\dag}(\p_0-ia_0)\psi - \frac{1}{2\mathcal{M}}\psi^{\dag}(\p_a-ia_a)^2\psi + ... \label{HLRaction}
\eeqn
where $\psi$ is non-relativistic, $\mathcal{M}$ is the effective mass. The RPA treatment takes $a_{\mu}$ as its mean field value, hence $\mathcal{L}_0$ processes zero Hall conductivity. The Fermi sea Berry $\Phi_{FS}$ is an extra ingredient to HLR theory, just as Berry phase is an extra input to Landau Fermi liquid theory. Suppose the Fermi sea carries $\Phi_{FS}$=$-2\pi\nu$ Berry phase, the induced anomalous Hall conductivity \cite{haldaneanomaloushall} $2\pi\sigma^{CF}_H$=$\frac{\Phi_{FS}}{2\pi}$ is then precisely what is needed to make $\sigma_H$ universal.

\subsubsection{Our action}
In the 1/2m action we proposed in the main text,
\beqn
\mathcal{L}_0 = & & i\bar\psi\gamma^{\mu}(\p_{\mu}-ia_{\mu})\psi - \frac{1}{2}\frac{1}{4\pi}ada,\nonumber\\
&-& \left(C_2-\frac{1}{2}+\frac{1}{2m}\right)\frac{1}{4\pi}ada + ...
\eeqn

Our reasoning on Berry phase in this case is a combination of that in HLR and Son's theory. With the extra prescription that Fermi sea has $-\left(\frac{1}{2m}-\frac{1}{2}\right)2\pi$ extra Berry phase, the anomalous Hall induced conductivity is $-\left(\frac{1}{2m}-\frac{1}{2}\right)$. Together with the mean field Hall conductivity, the total composite fermion Hall conductivity is then $2\pi\sigma^{CF}_H$=$-\left(\frac{1}{2m}-\frac{1}{2}\right) - \frac{1}{2} - \left(C_2-\frac{1}{2}+\frac{1}{2m}\right)$. Requiring $2\pi\sigma^{CF}_H$ to be $-\frac{1}{2m}$, we found $C_2$ has to be $\frac{1}{2}-\frac{1}{2m}$. Note that although Berry phase is defined modulo $2\pi$, $C_2$ cannot run, otherwise it cannot reduce to HLR action in the non-relativistic limit [as observed by Son \cite{Son}, Dirac fermion action reduces to HLR action in the non-relativistic limit]. Similar discussion works for $\nu=1-\frac{1}{2m}$ states.

\bibliography{onequarter.bib}

\end{document}